\documentclass[manuscript]{aastex}
\usepackage{longtable, enumerate, graphics, bm,graphicx, lscape, bm}

\begin{document}

\title{Recommended Thermal Rate Coefficients for the C + H$_3^+$ Reaction and Some Astrochemical Implications}
\author{S. Vissapragada\altaffilmark{1}, C. F. Buzard\altaffilmark{2}, K. A. Miller\altaffilmark{1}, A. P. O'Connor\altaffilmark{1, 3}, N. de Ruette\altaffilmark{1, 4}, \\ X. Urbain\altaffilmark{5}, and D. W. Savin\altaffilmark{1}}

\email{savin@astro.columbia.edu}
\altaffiltext{1}{\small Columbia Astrophysics Laboratory, Columbia University, New York, NY 10027, U. S. A.}
\altaffiltext{2}{\small Department of Chemistry, Barnard College, New York, NY 10027, U. S. A.}
\altaffiltext{3}{\small Present address: Max-Planck Institute for Nuclear Physics, Heidelberg 69117, Germany}
\altaffiltext{4}{\small Present address: Department of Physics, Stockholm University, Stockholm, \\ 106 91, Sweden}
\altaffiltext{5}{\small Institute of Condensed Matter and Nanosciences, Universit\'e catholique de Louvain, \\ B-1348 Louvain-la-Neuve, Belgium}
\begin{abstract}
We have incorporated our experimentally derived thermal rate coefficients for C + H$_3^+$ forming CH$^+$ and CH$_2^+$ into a commonly used astrochemical model. We find that the Arrhenius-Kooij equation typically used in chemical models does not accurately fit our data and use instead a more versatile fitting formula. At a temperature of 10 K and a density of 10$^4$ cm$^{-3}$, we find no significant differences in the predicted chemical abundances, but at higher temperatures of 50, 100, and 300 K we find up to factor of 2 changes. Additionally, we find that the relatively small error on our thermal rate coefficients, $\sim15\%$, significantly reduces the uncertainties on the predicted abundances compared to those obtained using the currently implemented Langevin rate coefficient with its estimated factor of 2 uncertainty.
\end{abstract}


\section{Introduction}
Interstellar astrochemistry is mostly organic in nature. Of the 194 molecules identified to date in the interstellar medium (ISM) and circumstellar shells, approximately three-quarters are carbon bearing \citep{m05}. Formation of these molecules begins with atomic carbon becoming bound into hydrocarbons \citep{vd98, h09}. This represents one of the first links in the chain of astrochemical reactions leading to the synthesis of complex organic molecules (COMs). A key reaction in this network is the proton transfer process \citep{w12}
\begin{equation}
\label{CH+}
\mathrm{C} + \mathrm{H_3^+} \rightarrow \mathrm{CH^+} + \mathrm{H_2}.
\end{equation}
In dense clouds, the resulting CH$^+$ is predicted to rapidly undergo sequential hydrogen abstraction with the abundant H$_2$ in the cloud to form CH$_3^+$, which has been identified as a bottleneck species in the chain of reactions leading to the formation of interstellar COMs \citep{ss95}. 

\par Current astrochemical models use the Langevin rate coefficient for Reaction (\ref{CH+}). However, our recent laboratory work has shown that the Langevin rate coefficient agrees poorly with the experimentally derived thermal rate coefficient \citep{o15}. Similarly, calculations by \citet{t91} and \citet{bc98, bc01}\textbf{,} using a combination of quantum mechanical potential energy surfaces and classical trajectories\textbf{,} do not agree with our experimental results for Reaction (\ref{CH+}). Additionally, we find that the reaction
\begin{equation}
\label{CH2+}
\mathrm{C} + \mathrm{H_3^+} \rightarrow \mathrm{CH_2^+} + \mathrm{H}
\end{equation}
is open, despite the lack of its inclusion in current astrochemical databases. We also find poor agreement between our results for Reaction (\ref{CH2+}) and semi-classical calculations. Moreover, our rate coefficient for this channel is larger than that of Reaction (\ref{CH+}) for temperatures below $\sim 50$~K. The resulting CH$_2^+$ then undergoes hydrogen abstraction to form CH$_3^+$.

\par \textbf{Reducing the uncertainty of the rate coefficient for Reaction~(\ref{CH+}) has been identified by \citet{w09, w10} as being critically important in order to more reliably predict the abundances for a large number of species observed in dense molecular clouds. Similarly, \citet{v08} has shown that the uncertainty in this rate coefficient hinders our ability to reliably predict chemical abundances in protoplanetary disks. Our recent laboratory studies have reduced the uncertainty on this reaction from a factor of 2 down to $\mathbf{\sim 15\%}$. Additionally, our work has demonstrated that Reaction~(\ref{CH2+}) is open at molecular cloud temperatures and should be included in the chemical network.}

\par \textbf{In this paper, we explore the astrochemical impact of our new rate coefficients for Reactions~(\ref{CH+}) and (\ref{CH2+}). First}, we have fit our experimentally derived thermal rate coefficients from \citet{o15} for \textbf{these reactions} to a simple functional form. Using our results, combined with the KInetic Database for Astrochemistry \citep[KIDA;][]{w15} and the gas-phase astrochemical code Nahoon \citep{w12}, we have \textbf{then} investigated the impact of our new data on astrochemical models. Below, we briefly discuss the \citet{o15} results in Section 2. We present the functional fits to the experimental data in Section 3. In Section 4, we briefly review the astrochemical model. Some astrochemical implications of our new thermal rate coefficients are discussed in Section 5, and a summary is presented in Section 6.

\section{Experimental Work}
\par \citet{o15} and \citet{d16}, respectively, measured reactions between H$_3^+$, with an internal energy of $\sim 2500$~K, and ground term atomic C and O, with the fine-structure levels statistically populated. In those works, we detail how we derived thermal rate coefficients from our data and discuss in detail their validity for astrochemical models.

\par To summarize those results for our carbon work, we found good agreement between our data and the mass-scaled results of \citet{s05}, who studied C with statistically populated fine-structure levels reacting with D$_3^+$ with an internal energy of 77~K.  Their work was carried out at a kinetic temperature of $\sim 1000$~K.  We found good agreement in the rate coefficients for both the CH$^+$ and CH$_2^+$ outgoing channels and for the sum of both channels.  Additionally, in our oxygen work, we found good agreement between the thermal rate coefficient summed over both the OH$^+$ and H$_2$O$^+$ outgoing channels compared to flow tube work at kinetic temperatures of $\approx 300$~K, which used H$_3^+$ with a corresponding level of internal temperature \citep{f76, mm00}. We refer the reader to \citet{o15} and \citet{d16} for more details.

\par Clearly, the optimal laboratory situation would be experimentally derived thermal rate coefficients involving thermally populated levels in C and H$_3^+$. However, such measurements appear to be beyond current experimental capabilities. Reliable calculations also appear to be just beyond current capabilities of quantum mechanical approaches. For now, the results of \citet{o15} represent the state of the art for Reactions (\ref{CH+}) and (\ref{CH2+}). For the rest of this paper, we follow the standard practice of extrapolating state-of-the-art laboratory results to the temperatures needed for molecular cloud studies.

\section{Fitting the Experimental Results}
\par
\begin{figure}
\centering
\resizebox{\textwidth}{!}{\includegraphics{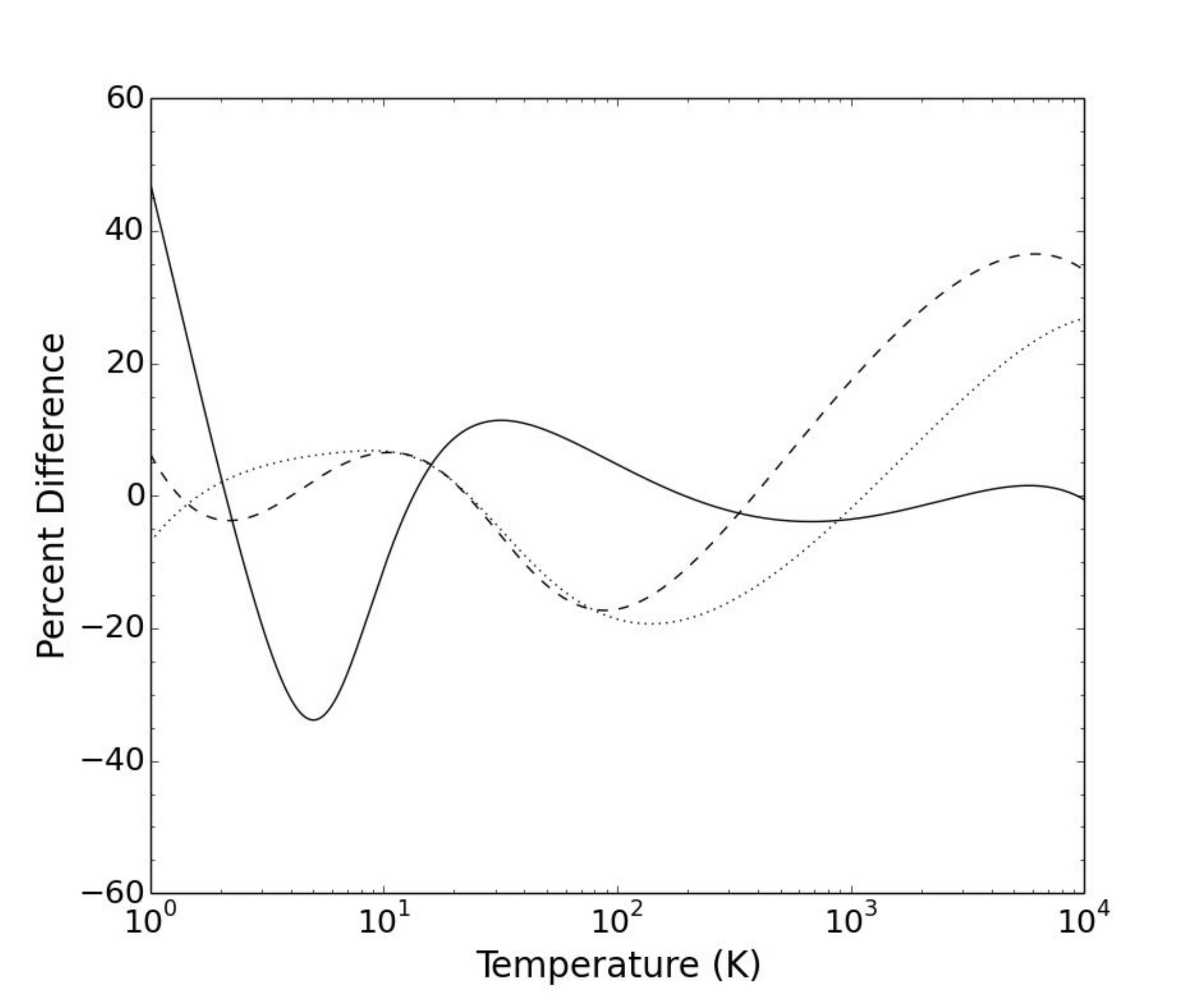}}
\caption{Percent difference between the fit to the experimentally derived thermal rate coefficients of \citet{o15} using \textbf{the Arrhenius-Kooij equation, given by} Equation (\ref{kooij})\textbf{,} and their actual data, normalized to their data. The solid curve represents the CH$^+$ formation channel, the dashed curve is for CH$_2^+$, and the dotted curve for the sum of these two channels.}
\label{AKfit}
\end{figure}

 Astrochemical databases typically store thermal rate coefficients using the Arrhenius-Kooij formula
\begin{equation}
\label{kooij}
k(T) = A\bigg(\frac{300\ \mathrm{K}}{T}\bigg)^{-B}\textup{exp}\bigg(\frac{-C}{T}\bigg).
\end{equation}
However, our experimentally derived thermal rate coefficients for Reactions (\ref{CH+}) and (\ref{CH2+}) cannot be accurately reproduced by this formula. This can be seen in Figure \ref{AKfit}, which presents the fit using Equation (\ref{kooij}) to the \citet{o15} data minus their actual data, normalized by their results (i.e., the normalized residuals). These findings continue the trend seen in merged-beams astrochemical studies that the Arrhenius-Kooij formula does a poor job in reproducing the experimentally derived rate coefficients for barrierless exoergic reactions, as has been seen for associative detachment \citep{k10}, dissociative recombination \citep{n13, n14}, and proton and H$_2^+$ transfer reactions \citep{d16}.
\par \textbf{In the absence of any deep theoretical understanding of Reactions (\ref{CH+}) and (\ref{CH2+}), it is not clear what fitting formula to use. We have opted to use the versatile fitting function recommended by \citet{n13} for astrochemical modeling, namely,}
\begin{equation}
\label{novotny}
k(T) = A\bigg(\frac{300\ \mathrm{K}}{T}\bigg)^n + T^{-3/2} \sum\limits_{i=1}^\mathbf{4} c_i \textup{exp}\bigg(\frac{-T_i}{T}\bigg).
\end{equation}
\textbf{We have used this equation to fit the thermal rate coefficient data of \citet{o15} for Reactions (\ref{CH+}) and (\ref{CH2+}) to better than 1\% over the $\mathbf{1-10,000}$~K temperature range.} The relevant fitting parameters are listed in Table \ref{coefdata}. In that table, we also give a fitted rate coefficient for the sum of both channels.

\section{Astrochemical Model}
\par In order to study the impact of the experimentally derived thermal rate coefficients from \citet{o15} on astrochemical models of dark molecular clouds, we have used the Nahoon code along with the kida.uva.2014 astrochemical database \citep{w12, w15}. The database currently contains 489 species and 7509 reactions. These include Reaction (\ref{CH+}) but not Reaction (\ref{CH2+}). We have modified KIDA so that we can run the model using either our fitted rate coefficient data for both these channels or for the sum of these two channels.

\par The input parameters used were typical values for dark molecular clouds \citep{n00, rf10}. For each run, the cosmic ray ionization rate $\zeta$ was taken to be $10^{-17}$ $\mathrm{s^{-1}}$ and the visual extinction $A_v$ was set to $30$. The initial chemical abundances were taken from \citet{w15}, \textbf{and are reproduced in Table~\ref{initial}}. Each simulation used a fixed cloud temperature, $T$, between 10 to 300~K and a fixed total number density of hydrogen nuclei, $n_\mathrm{H}$, in a range from $10^3$ to $10^7$ cm$^{-3}$.

\section{Astrochemical Implications}
\par We first look at how the differences in the temperature dependence and magnitude of our rate coefficients relative to those of the Langevin rate coefficient affect predicted chemical abundances. Next, we investigate how the $\sim 15\%$ error reported in \citet{o15} reduces the uncertainty in the predicted abundances compared to the uncertainties resulting from the estimated factor of 2 error in the Langevin value.

\par In order to test the sensitivity of our results to the experimentally derived branching ratio for forming CH$^+$ or CH$_2^+$, we have run the model treating both reactions separately and with both reactions summed together. For the results presented below, we find no significant difference in the model output for either assumption. We attribute this to the high H$_2$ abundance in the dark clouds, resulting in rapid hydrogen abstraction reactions.

\subsection{Predicted Abundances}
\par
\begin{landscape}
\begin{figure}
\centering
\resizebox{650pt}{!}{\includegraphics{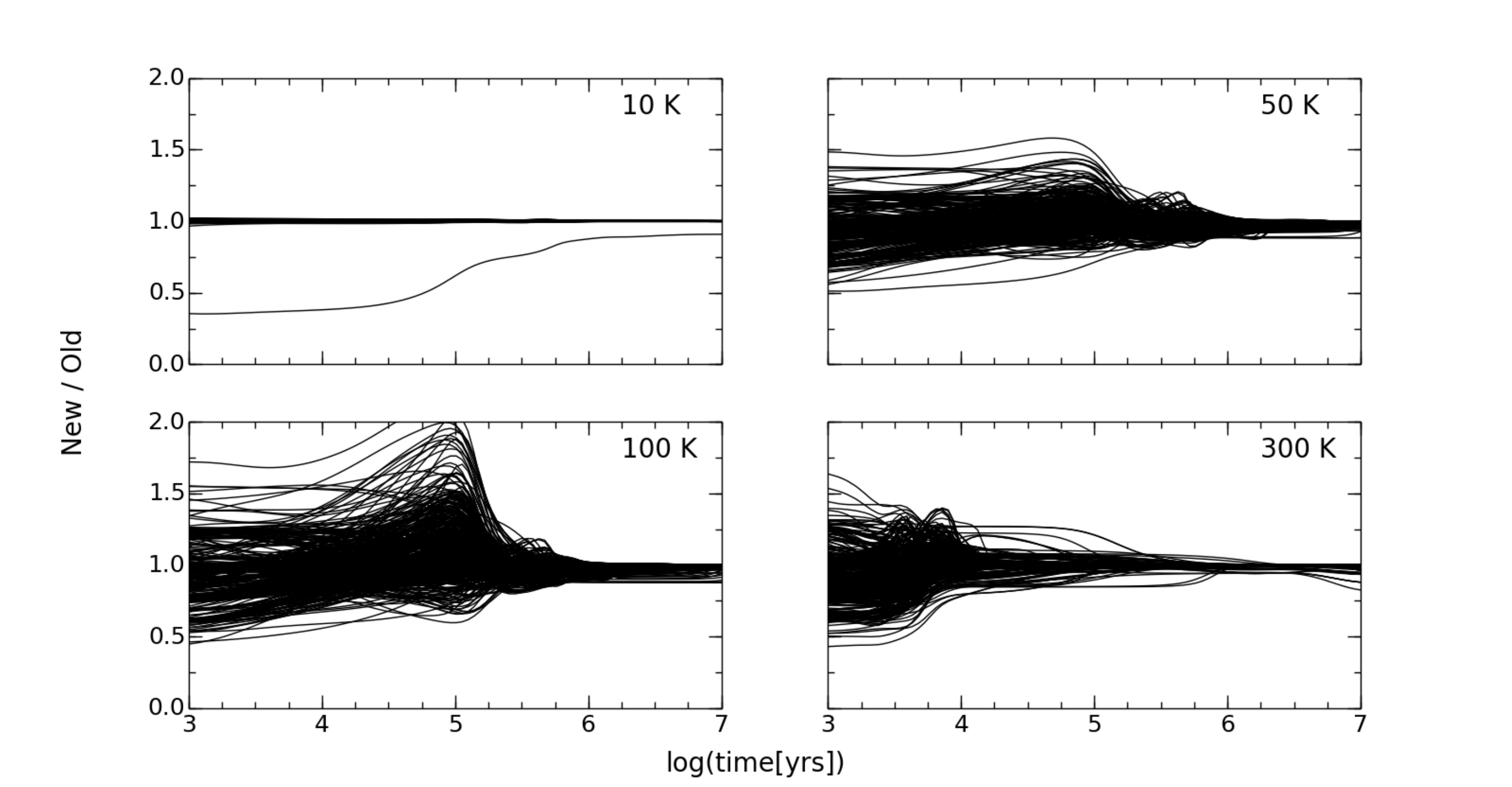}}
\caption{Ratio of the predicted abundances for all 489 species in KIDA using our new rate coefficients divided by those using the unmodified model (old) rate coefficients. Results are shown at $n_\mathrm{H}$~=~10$^4$~cm$^{-3}$ for the temperatures indicated on each plot.}
\label{abundances}
\end{figure}
\end{landscape}

 Figure \ref{abundances} shows the fractional difference in the predicted abundances for all 489 species using our new rate coefficients relative to those from the unmodified model. Specifically, we have plotted the new abundances normalized by the old abundances. Calculations were carried out for $n_\mathrm{H}$ = 10$^4$ cm$^{-3}$ and $T$ = 10, 50, 100, and 300~K.

\par At 10 K, we find no significant differences except for CH$^+$. The abundances of all other species are essentially unchanged because, as noted by \citet{o15}, any CH$^+$ and CH$_2^+$ formed rapidly undergo hydrogen abstraction, leading to CH$_3^+$, and the summed rate coefficient at 10 K for Reactions (\ref{CH+}) and (\ref{CH2+}) is basically equal to the Langevin value currently used in the database for Reaction (\ref{CH+}). Hence the absence of Reaction (\ref{CH2+}) in the databases appears not to be an issue at this temperature. The decreased abundance for CH$^+$ in the new model is due to the decreased rate coefficient for Reaction (\ref{CH+}). Naively, one would expect the CH$_2^+$ abundance to increase due to the addition of Reaction (\ref{CH2+}) to the network. However, this is compensated for by a reduction in the hydrogen abstraction rate for CH$^+$ forming CH$_2^+$ due to the decreased CH$^+$ abundance. Hence the CH$_2^+$ abundance remains unchanged.

\par At 50 K, our summed rate coefficient is a factor of $\sim 30\%$ smaller than the Langevin value. For species that depend on the products of Reactions (\ref{CH+}) and (\ref{CH2+}), this means their abundances decrease in the modified model using our data. Conversely, for species whose formation depends on C and H$_3^+$, the abundance of those species increases with the new model as C and H$_3^+$ are destroyed less rapidly using our new rate coefficients. The new predicted abundances range from a factor of 2 smaller than the old to a factor of 1.5 larger; however, this spread decreases dramatically between $10^5$ and $10^6$~years. This appears to be due to a large increase in the abundance of O$_2$ during this epoch, which enables the reaction
\begin{equation}
\label{cdestr1}
\mathrm{C} + \mathrm{O_2} \rightarrow \mathrm{O} + \mathrm{CO}
\end{equation}
to become important. This leads to a dramatic decrease in the atomic C abundance, thereby reducing the importance of reactions (\ref{CH+}) and (\ref{CH2+}).

\par At 100 K, the summed rate coefficient is $\sim 40\%$ smaller than the Langevin value, and this leads to correspondingly larger variations in the predicted abundances plotted in Figure \ref{abundances}. The new abundances range from a factor of 2 smaller to a factor of 2 larger. As before, these variations decrease dramatically between $10^5$ and $10^6$~years. This is again due to a large increase in the O$_2$ abundance, an increase in the importance of Reaction (\ref{cdestr1}), and an accompanying decrease in the C abundance.

\par At 300 K, the total rate coefficient is $\sim 45\%$ smaller than the Langevin value, leading to new abundances which range from a factor of 2 smaller than the old to a factor of 1.5 larger. Also, at this higher temperature a new set of chemical reactions become important, leading to a dramatic decrease in the atomic C abundance at around $10^{3.5}$~years. This decrease appears to be due, in large part, to an increase in the abundance of neutral hydrocarbons, which react with and incorporate much of the atomic C in the cloud. As a result of the decreased atomic C abundance, at this temperature, Reactions (\ref{CH+}) and (\ref{CH2+}) are less important for cloud ages above $10^{3.5}$~years.

\subsection{Abundance Uncertainties}
\par Langevin rate coefficients have estimated uncertainties of a factor of 2, though our previous work indicates that the actual uncertainties in Langevin rate coefficients may be even larger \citep{k10, o15, d16}. \citet{o15} report uncertainty factors of $\approx \pm 13\%$ and $\approx \pm 18\%$ for their experimentally derived thermal rate coefficients for Reactions (\ref{CH+}) and (\ref{CH2+}), respectively. To track the resulting decrease of uncertainty throughout the network, we first ran the model using the Langevin rate coefficient at the upper limit of the estimated factor of 2 uncertainty, and then at the lower limit of its uncertainty. For each species, the abundances from those runs ($\chi_\mathrm{upper}$ and $\chi_\mathrm{lower}$, respectively) were obtained as a function of time. The difference of the logarithms of the two abundances, $\log(\chi_\mathrm{upper}/\chi_\mathrm{lower})$, was used as a heuristic for the level of uncertainty in the predicted abundances. Then, we replaced the Langevin value with our new coefficients and ran the model again using the new upper and lower uncertainty limits. By tracking the uncertainty statistic $\log(\chi_\mathrm{upper}/\chi_\mathrm{lower})$ for the old and new models, we were able to track the reduction in the abundance uncertainties throughout the network.

\par
\begin{figure}
\centering
\resizebox{\textwidth}{!}{\includegraphics{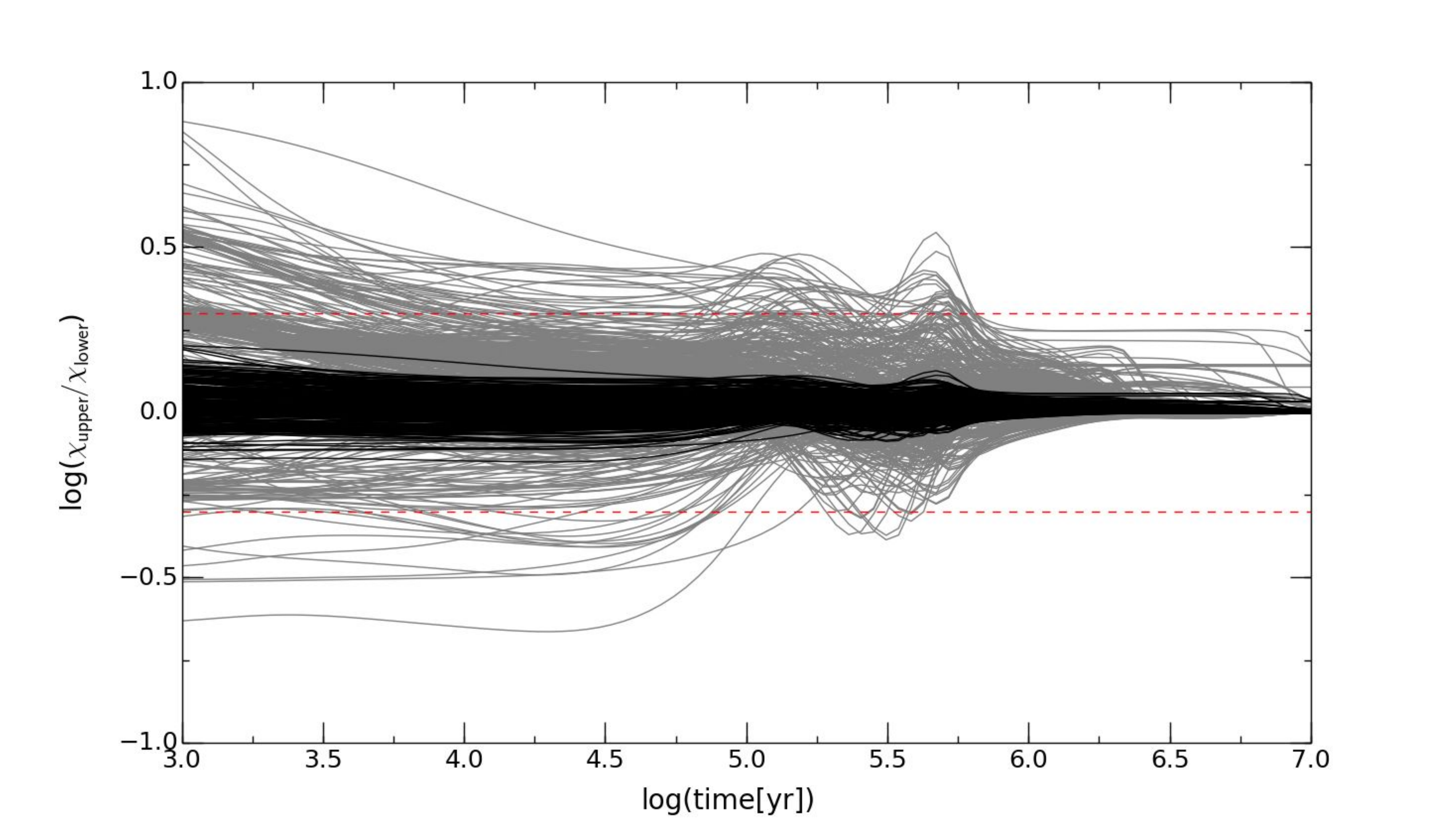}}
\caption{Uncertainty plot for $T = 10$ K and $n_\mathrm{H} = 10^4$ $\mathrm{cm^{-3}}$, in which the uncertainty statistic, $\log(\chi_{\mathrm{upper}}/\chi_{\mathrm{lower}})$, is expressed as a function of time. The dashed red lines represent the thresholds for ``significant" uncertainty as defined by \citet{w15}. The solid lines represent the uncertainty of individual species in KIDA. The lighter curves show the old uncertainties of all species in KIDA due to Reaction (\ref{CH+}), while the darker curves show the updated uncertainties of those species due to our new data for Reactions (\ref{CH+}) and (\ref{CH2+}).} 
\label{uncertainty1}
\end{figure}

\begin{figure}
\centering
\resizebox{\textwidth}{!}{\includegraphics{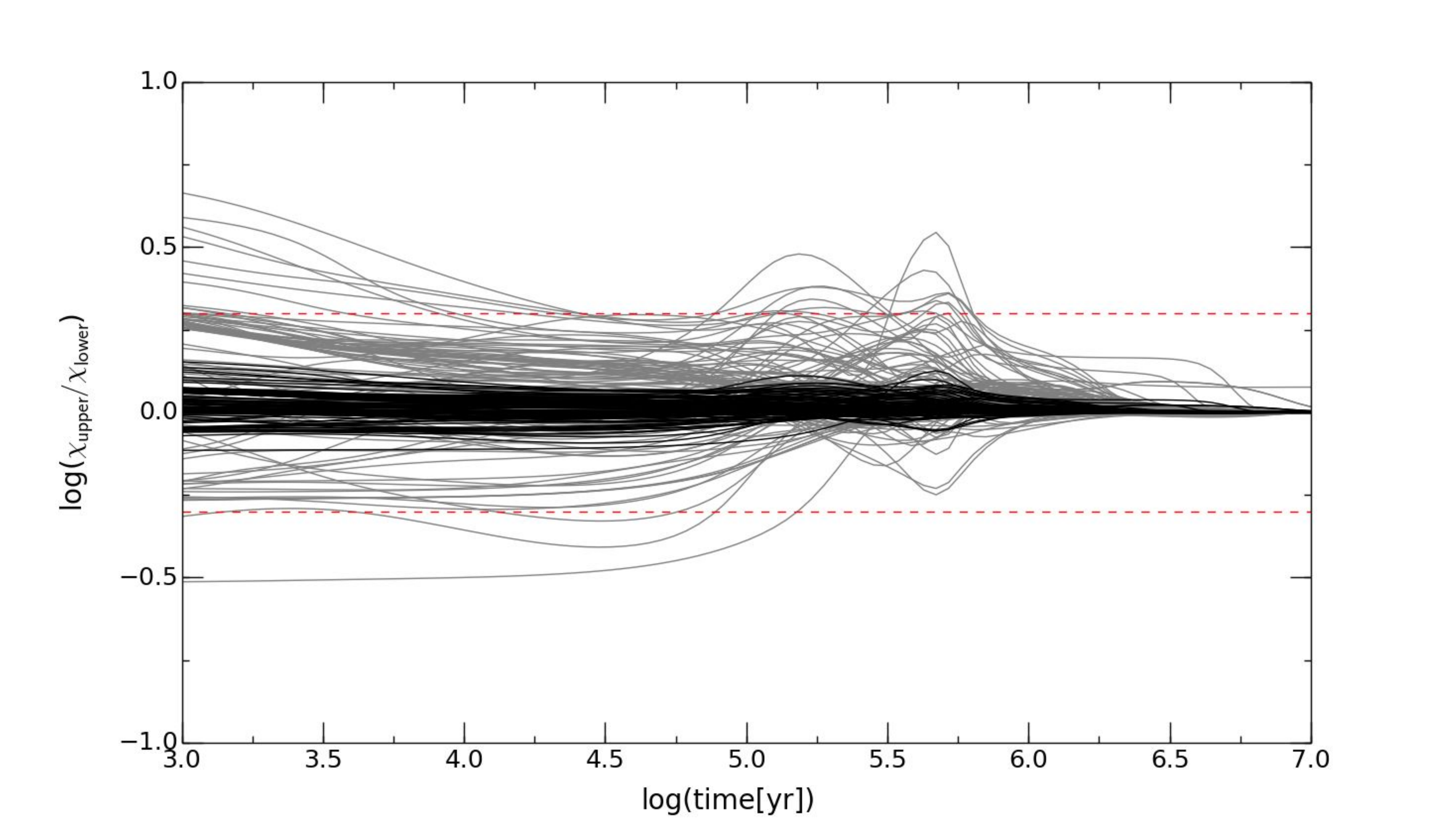}}
\caption{Same as Figure 3, but with cuts made for species observed in the ISM \citep{m05}.} 
\label{uncertainty2}
\end{figure}

\begin{landscape}
\begin{figure}
\centering
\resizebox{650pt}{!}{\includegraphics{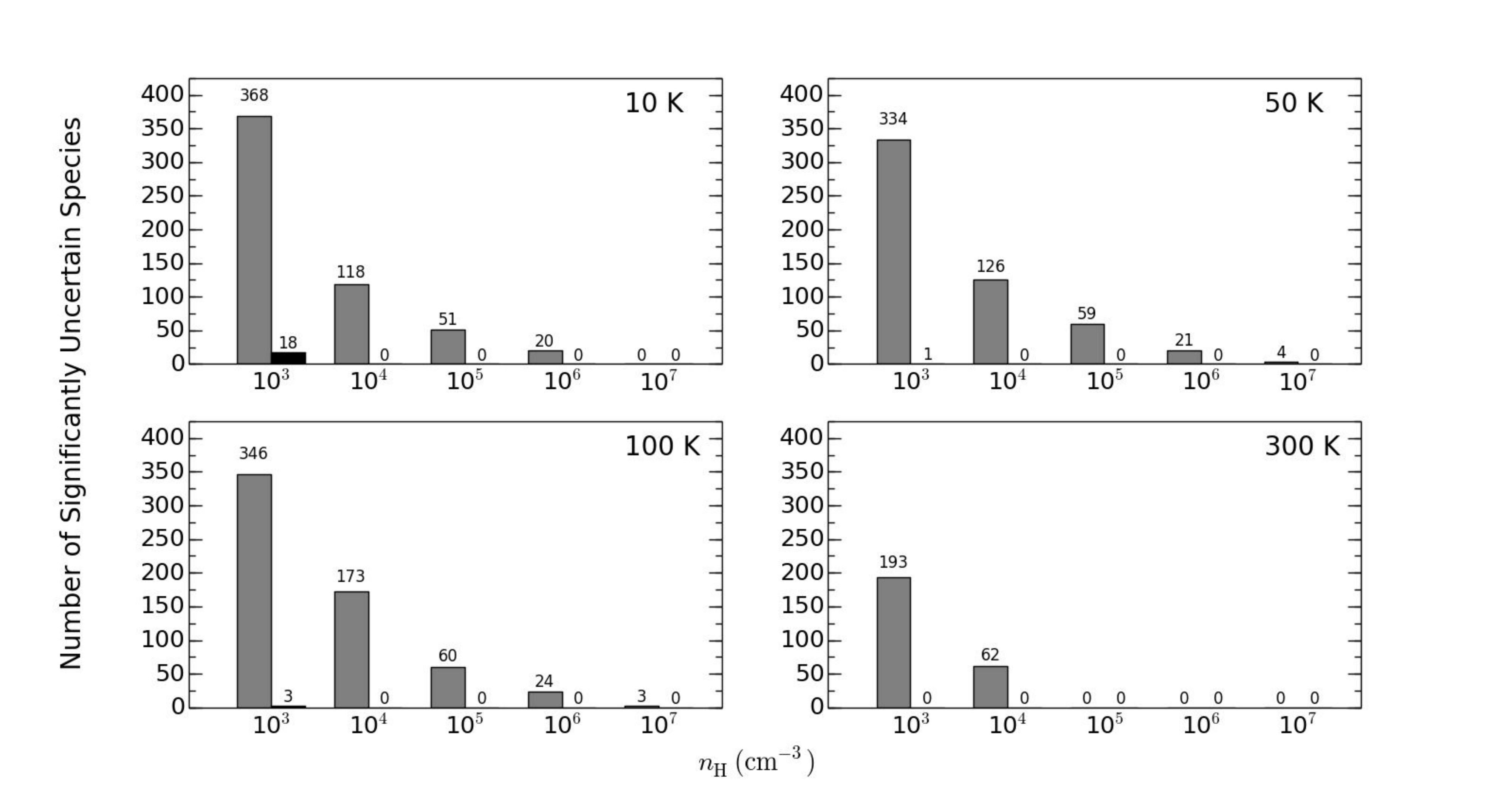}}
\caption{Number of significantly uncertain species as a function of density for the temperatures indicated on each plot. The lighter bars show the number of significantly uncertain species for the old model, while the darker bars show that for the new model.}
\label{unctempdens}
\end{figure}
\end{landscape}

 Following \citet{w15}, a ``significantly" uncertain species was taken to mean that $|\log(\chi_\mathrm{upper}/\chi_\mathrm{lower})|$ was greater than or equal to 0.3 (i.e., a factor of 2 difference). We find that for every temperature and density in our model, there was a reduction in the number of significantly uncertain species. Figure \ref{uncertainty1} shows this for 10~K and $10^4$~cm$^{-3}$ by overplotting the uncertainty statistic for the old and new networks. In this figure, 118 species in the old model are shown to be significantly uncertain, while no species in the new models are. Figure \ref{uncertainty2} does the same, but only for species that have been observed in the ISM. In this figure, 17 species in the old model are significantly uncertain. These species (listed here by number of atoms) are NaOH, SO$_2$, HOOH, HCOOH, HC$_4$N, CH$_3$CN, C$_2$H$_4$, CH$_3$CHO, CH$_2$CHCN, HCOOCH$_3$, CH$_3$CH$_2$OH, CH$_3$OCH$_3$, CH$_3$C$_4$H, CH$_3$COCH$_3$, CH$_3$C$_5$N, CH$_3$C$_6$H, and C$_6$H$_6$. No species in the new model are significantly uncertain. Additionally, Figure \ref{unctempdens} shows the relationship between the number of significantly uncertain species and $n_\mathrm{H}$ for $T$ = 10, 50, 100, and 300~K. The number of significantly uncertain species has been reduced over the full range of temperatures and densities.

\section{Summary}
\par In this work, we have fit the experimentally derived thermal rate coefficients of \citet{o15} for Reactions (\ref{CH+}) and (\ref{CH2+}) to a functional form given by Equation (\ref{novotny}), as the Arrhenius-Kooij formula gives a poor fit to the data. We then included these results into the KIDA/Nahoon astrochemical model to determine their impact on the predicted abundances and their uncertainties as a function of temperature and density. At 10 K, the summed rate coefficient of \citet{o15} matches the Langevin value, leading to no significant difference between the old and new models for predicted abundances. However, at higher temperatures the updated rate coefficient decreases relative to the Langevin value, leading to significant differences in predicted abundances. Additionally, we show that the smaller uncertainties on the \citet{o15} data, compared to that of the Langevin value, lead to a reduction in the uncertainties of the predicted abundances.

\acknowledgments
The authors thank \textbf{R. G. Bhaskar} and V. Wakelam for stimulating conversations. \textbf{We also thank the anonymous referee for their valuable comments.} This work was supported in part by the NSF Division of Astronomical Sciences Astronomy and Astrophysics Grants Program and by the NASA Astronomy and Physics Research and Analysis Program. X.U. is Senior Research Associate of the FRS-FNRS.  S.V. was supported in part by the NSF Research Experience for Undergraduates program.

\clearpage

\clearpage
\clearpage

\clearpage

\clearpage

\clearpage
\begin{deluxetable}{@{\extracolsep{4pt}} c r l r l r l l@{}}
\tablewidth{0pt}
\tablecolumns{5} 
\tablecaption{Fit parameters using Equation (\ref{novotny}) for the thermal rate coefficient of C + H$_3^+$ forming either CH$^+$ or CH$_2^+$. Also given is the rate coefficient summed over both channels.}
\tablehead{ 
\colhead{Parameter} & 
\multicolumn{2}{c}{CH$^+$} & 
\multicolumn{2}{c}{CH$_2^+$}  &
\multicolumn{2}{c}{Sum} & 
\colhead{Units}		\\
\cline{2-3} \cline{4-5} \cline{6-7}
 \colhead{} & \colhead{$x$}& \colhead{$y$} & \colhead{$x$}& \colhead{$y$} & \colhead{$x$}& \colhead{$y$} & \colhead{}
}

\startdata
$A$		& \phs $\mathbf{6.93}$	& $\mathbf{-10}$	&  \phn $\mathbf{3.35}$	& $\mathbf{-10}$	&  $\mathbf{1.04}$	& $\mathbf{-9}$		& ${\rm cm^3\ s^{-1}}$ \\
$n$		& $\mathbf{-8.34}$		& $\mathbf{-2}$ 	&  \phn $\mathbf{1.89}$	& $\mathbf{-1}$		&  $\mathbf{2.31}$	& $\mathbf{-3}$		& dimensionless\\
$c_1$	& $\mathbf{-7.24}$		& $\mathbf{-9}$ 	&  \phn $\mathbf{2.73}$	& $\mathbf{-8}$		&  $\mathbf{3.40}$	& $\mathbf{-8}$		& ${\rm K^{3/2}\ cm^3\ s^{-1}}$ \\
$c_2$	& $\mathbf{-9.07}$ 		& $\mathbf{-10}$ 	&  \phn $\mathbf{5.58}$	& $\mathbf{-9}$		&  $\mathbf{6.97}$	& $\mathbf{-9}$		& ${\rm K^{3/2}\ cm^3\ s^{-1}}$ \\
$c_3$	& \phs $\mathbf{7.48}$	& $\mathbf{-8}$ 	&  \phn $\mathbf{7.46}$	& $\mathbf{-8}$		&  $\mathbf{1.31}$	& $\mathbf{-7}$		& ${\rm K^{3/2}\ cm^3\ s^{-1}}$ \\
$c_4$	& \phs $\mathbf{9.93}$	& $\mathbf{-5}$ 	&  $\mathbf{-1.92}$		& $\mathbf{-4}$		&  $\mathbf{1.51}$	& $\mathbf{-4}$		& ${\rm K^{3/2}\ cm^3\ s^{-1}}$ \\
$T_1$	& \phs $\mathbf{8.01}$	& \phs $\mathbf{0}$	&  \phn $\mathbf{6.49}$	& \phs $\mathbf 0$	&  $\mathbf{7.62}$	& \phs $\mathbf{0}$		& K \\
$T_2$	& \phs $\mathbf{1.92}$	& \phs $\mathbf 0$ 	&  \phn $\mathbf{1.30}$	& \phs $\mathbf 0$	&  $\mathbf{1.38}$	& \phs $\mathbf{0}$		& K \\
$T_3$	& \phs $\mathbf{4.19}$	& \phs $\mathbf 1$ 	&  \phn $\mathbf{1.90}$	& \phs $\mathbf 1$	&  $\mathbf{2.66}$	& \phs $\mathbf{1}$		& K \\
$T_4$	& \phs $\mathbf{8.08}$	& \phs $\mathbf 3$ 	&  \phn $\mathbf{1.62}$	& \phs $\mathbf 4$	&  $\mathbf{8.11}$	& \phs $\mathbf{3}$		& K \\
\enddata
\tablecomments{The value for each parameter is given by $x \times 10^{y}$.}
\label{coefdata}
\end{deluxetable}

\clearpage
\begin{deluxetable}{ c  c  c}
\tablewidth{0pt}
\tablecolumns{3} 
\tablecaption{\textbf{Initial chemical abundances with respect to H nuclei, with the value for each abundance given by $\bm x \mathbf{\times 10}^{\bm y}$.}}
\tablehead{ 
\colhead{\textbf{Species}} & 
\multicolumn{2}{c}{\textbf{Abundance}} \\
\cline{2-3}
\colhead{} & \colhead{$\bm x$} & \colhead{$\bm{y}$}
}

\startdata
\textbf{He}& \textbf{9.0}		& \textbf{-2\phn} \\
\textbf{C}	& \textbf{1.7}	 	& \textbf{-4\phn} \\
\textbf{S}	& \textbf{8.0}		& \textbf{-8\phn} \\
\textbf{Fe}	& \textbf{3.0}		& \textbf{-9\phn} \\
\textbf{Mg}& \textbf{7.0}		& \textbf{-9\phn} \\
\textbf{P}	& \textbf{2.0}		& \textbf{-10} \\
\textbf{N}	& \textbf{6.2}		& \textbf{-5\phn} \\
\textbf{O}	& \textbf{2.4}		& \textbf{-4\phn} \\
\textbf{Si}	& \textbf{8.0}		& \textbf{-9\phn} \\
\textbf{Na}	& \textbf{2.0}		& \textbf{-9\phn} \\
\textbf{Cl}	& \textbf{1.0}		& \textbf{-9\phn}
\enddata
\label{initial}
\end{deluxetable}

\end{document}